\documentclass[journal]{IEEEtran}
\IEEEoverridecommandlockouts
\usepackage{cite}
\usepackage{amsmath,amssymb,amsfonts}
\usepackage{graphicx}
\usepackage{textcomp}
\usepackage{xcolor}
\usepackage{stfloats}
\usepackage[0]{editing}
\usepackage{enumitem}
\usepackage{algorithmic}
\usepackage{siunitx}
\usepackage{subfig}
\usepackage{hyperref}
\usepackage{fancyhdr}
\usepackage{lastpage}
\usepackage{hyperref}
\hypersetup{
    colorlinks=true,
    linkcolor=blue,
    citecolor=blue,
    filecolor=blue,
    urlcolor=blue
}

\usepackage[linesnumbered,ruled,vlined]{algorithm2e}

\def\BibTeX{{\rm B\kern-.05em{\sc i\kern-.025em b}\kern-.08em
    T\kern-.1667em\lower.7ex\hbox{E}\kern-.125emX}}
    
\begin{document}

\title{\add{Pilot and Data Power Control for Uplink Cell-free  massive MIMO}\\
{\footnotesize }}

\author{Saeed~Mohammadzadeh$^{\dag}$,  Mostafa~Rahmani$^{\dag}$,  Kanapathippillai~Cumanan$^{\dag}$, 
Alister~Burr$^{\dag}$,
 and~ Pei~Xiao$^{\dag\dag}$, 
 \thanks{$^\dag$Department of Electronic Engineering, University of York, UK.  $^{\dag\dag}$ Institute for Communication Systems, University of Surrey, UK.
} \vspace{-0.75em}}

\maketitle

\begin{abstract}

\add{This paper introduces a novel iterative algorithm for optimizing pilot and data power control (PC) in cell-free massive multiple-input multiple-output (CF-mMIMO) systems, aiming to enhance system performance under real-time channel conditions. The approach begins by deriving the signal-to-interference-plus-noise ratio (SINR) using a matched filtering receiver and formulating a min-max optimization problem to minimize the normalized mean square error (NMSE). Utilizing McCormick relaxation, the algorithm adjusts pilot power dynamically, ensuring efficient channel estimation. A subsequent max-min optimization problem allocates data power, balancing fairness and efficiency. The iterative process refines pilot and data power allocations based on updated channel state information (CSI) and NMSE results, optimizing spectral efficiency. By leveraging geometric programming (GP) for data power allocation, the proposed method achieves a robust trade-off between simplicity and performance, significantly improving system capacity and fairness. The simulation results demonstrate that dynamic adjustment of both pilot and data PC substantially enhances overall spectral efficiency and fairness, outperforming the existing schemes in the literature. }


\end{abstract}


\section{Introduction}

Mobile user equipment (UEs) will keep growing with increasing capacity demands over the next decade, requiring ubiquitous and boundless connectivity. Massive multiple-input multiple-output (mMIMO) is widely recognized as one of the most promising techniques for the 5G and beyond wireless networks, offering significant improvements in spectral efficiency (SE) along with simplified processing and near-optimal performance \cite{marzetta2010noncooperative,larsson2014massive,parkvall2017nr}. Looking ahead, the main constraints on spectral efficiency (SE) are inter-cell interference due to lack of access point (AP) cooperation, significant path losses, and corresponding hardware energy consumption \cite{buzzi2016survey}. Sixth-generation (6G) networks aim to enhance SE by 100 times compared to 5G by addressing these issues with denser, cell-free network infrastructure,  transitioning from a cell-centric to a user-centric approach to guarantee comprehensive coverage and high data rate \cite{saad2019vision}.

Unlike traditional cellular networks, CF-mMIMO employs numerous distributed APs equipped with multiple antennas to serve multiple UEs simultaneously, resulting in significant improvements in coverage, capacity, and \add{reliability \cite{ngo2013energy,bashar2019uplink,bashar2020performance,rahmani2022deep,chu2023joint,chong2024performance,chong2024distribution,chong2024statistical}}.
However, one of the key challenges in this system is the efficient allocation of pilot and data power, which play a crucial role in establishing reliable and high-quality communication links between APs and UEs while alleviating pilot contamination. It is known that the pilot power controls the channel estimation quality \cite{rahmani2022multi}, while the data power determines the spectral efficiency and user data rates. Hence, it is vital to optimize power allocation strategies to improve system performance and meet quality-of-service (QoS) requirements for UEs while the energy consumption is reduced in the central processing unit (CPU) \cite{bjornson2014massive}. \\
\indent In CF-mMIMO systems, joint pilot and data power control (PC) is a method that simultaneously manages the power levels of both pilot and data signals to optimize overall system performance. By jointly optimizing the allocation of pilot and data powers, we can effectively leverage the spatial diversity of the wireless channel to mitigate key challenges such as pilot contamination, inter-user interference, and noise. Specifically, data power control plays a crucial role in managing inter-user interference, reducing the overlapping interference between users during data transmission. Meanwhile, pilot power control is instrumental in addressing the effects of pilot contamination, enhancing channel estimation accuracy. This dual approach, balancing data and pilot power, enables a more robust and interference-resilient communication environment in CF-mMIMO systems.
This problem is challenging to solve because the data power and pilot power optimization problems are non-convex when addressed individually. When combined, they create an even more complex, non-convex optimization problem. Additionally, these two vectors—data power and pilot power—are interdependent, adding another layer of complexity. This interdependence requires careful balancing to achieve an optimal solution that maximizes spectral efficiency while mitigating interference and pilot contamination.

\indent The works in \cite{cheng2016optimal,ghazanfari2018optimized,van2020power} investigated the effects of the pilot and data PC in the uplink of cell-based systems. However, previous optimization algorithms used in cellular systems cannot be applied to cell-free systems. Each serving AP needs to estimate its wireless channel to its UEs based on the same uplink pilot signal in cell-free systems. On the other hand, some research on cell-free systems focused on optimizing only the data transmission powers; there is no pilot power control \cite{ngo2017cell}, \cite{nguyen2018optimal} while using full and fixed power to transmit pilot signals.\\
\indent In \cite{mai2018pilot}, an algorithm was proposed based on a pilot PC scheme to minimize the maximum channel estimation error among users. The algorithm used the data power allocation scheme proposed in \cite{ngo2017cell} to solve a max-min fairness problem, while the AP selection is performed based on the largest large-scale fading-based criteria. However, the power levels for both pilot and data signals are not optimized together, meaning their adjustments are not determined simultaneously. The joint PC method was considered in \cite{masoumi2018joint} by utilizing single-antenna APs wherein the UEs are subject to a limited total energy. The study in \cite{mai2020design} proposed a method utilizing first-order optimization to adjust pilot and data power based on a greedy power algorithm. Although the authors describe this approach as joint, it primarily involves sequential optimization of pilot and data power levels rather than fully integrating them. The authors in \cite{braga2021joint} considered the joint pilot and data PC in user-centric scenarios by combining successive convex approximation and geometric programming (GP) and further extended the idea based on deep reinforcement learning in \cite{braga2022decentralized}. In \cite{liu2022joint}, a sequential convex approximation algorithm is utilized to solve GP to control the pilot and data power jointly.

\add{This study presents an algorithm that iteratively controls both pilot and data power levels to enhance the performance of a wireless communication system. First, we derive each user's signal-to-interference-plus-noise ratio (SINR) based on a matched filtering receiver. Then, we establish the normalized mean square error (NMSE) and formulate a min-max optimization problem to minimize it. This optimization problem involves additional novel non-convex constraints, which we address using McCormick relaxation techniques, allowing pilot power levels to adapt to real-time channel conditions.
The algorithm then solves a max-min optimization problem to determine optimal data power levels, enhancing overall power control for effective data transmission. The key innovation in this approach lies in its iterative nature: pilot power levels are adjusted based on the resulting data power allocations, and data power levels are subsequently recalculated in response to the updated pilot power settings. This cycle of adjustments continues, with each iteration refining the pilot and data power vectors in response to the latest channel state information (CSI) and NMSE results. This iterative approach ensures that both pilot and data power settings evolve to optimize channel estimation accuracy and data transmission quality for each user. By tailoring power levels according to CSI and NMSE, users with poor channel estimates receive higher power to improve performance, while those with strong channel conditions are assigned lower power to reduce interference and conserve energy. Leveraging GP techniques \cite{chiang2007power,boyd2007tutorial}, we allocate data power based on the max-min fairness optimization problem, achieving a balanced approach that improves overall system efficiency and user experience.\\
\indent Unlike \cite{mai2018pilot,liu2022joint}, which focuses solely on pilot and data power control in separate optimization problems and then uses it for the spectral efficiency calculation, our method integrates pilot power and data power control dynamically and iteratively, ensuring that the users with poor channel information receive adequate pilot power without compromising overall data transmission power efficiency. Through extensive simulations, we demonstrate the effectiveness and efficiency of the proposed scheme in improving system performance and achieving optimal power allocation in CF-mMIMO deployments. The paper's contributions and results are outlined as follows:}
\add{\begin{itemize}
    \item To enhance the SE of the CF-mMIMO system, we propose a novel algorithm based on the power allocation scheme for the uplink, which is designed solely based on the channel's statistics. 
    \item We derive an algorithm based on NMSE minimization using the channel statistic and the effects of channel estimation errors. We introduce a novel constraint based on the maximum transmit budget at the users and the coherence block. Then, we employ the McCormick relaxation technique to adjust pilot power levels based on real-time channel conditions, reducing interference and improving the accuracy of CSI estimation.
    \item  By tailoring pilot power allocation to each user’s NMSE, the system dynamically adapts to variations in channel conditions across users, and the data power is changed adaptively. This approach enhances system efficiency by balancing power, minimizing interference, and improving spectral efficiency.
    \item We provide a complexity analysis of the proposed and the compared schemes and present numerical results that validate the theoretical derivations of the proposed scheme's optimality. 
\end{itemize}}
\subsection{Outline}    
The rest of the paper is organized as follows: Section II outlines the system model. In Section III, we propose an iterative-based PC algorithm that optimizes both pilot and data power allocation to enhance SE. The section also includes the summary and the complexity of the proposed method. Section IV presents simulation results, highlighting the benefits of the proposed method. Finally, Section V concludes the paper.
\subsection{Notations}
We represent vectors using bold lowercase letters and matrices using bold uppercase letters. $\mathbb{E} \{\cdot\}$ stands for the statistical expectation of random variables, and a circular symmetric complex Gaussian matrix with covariance $\mathbf{Z}$ is denoted by $\mathcal{CN}(0,\mathbf{Z})$. The diagonal matrix is denoted as $\text{diag}(\cdot)$, and the symbol $\mathbb{C}$ is used for the complex numbers. $(\cdot)^\mathrm{T}$, $(\cdot)^\mathrm{*}$, $(\cdot)^{-1}$ and $(\cdot)^\mathrm{H}$ denote transpose, conjugate, inverse, and conjugate-transpose, respectively. The Euclidean norm and the absolute value are represented by $\|\cdot \|$ and $|\cdot |$, respectively. We use the superscripts $^\mathrm{p}$ and $^\mathrm{d}$ to represent the variables or parameters associated with pilot and data. 
\begin{figure}[t]
	\centering
	\includegraphics[width=2.7in,height=1.9in]{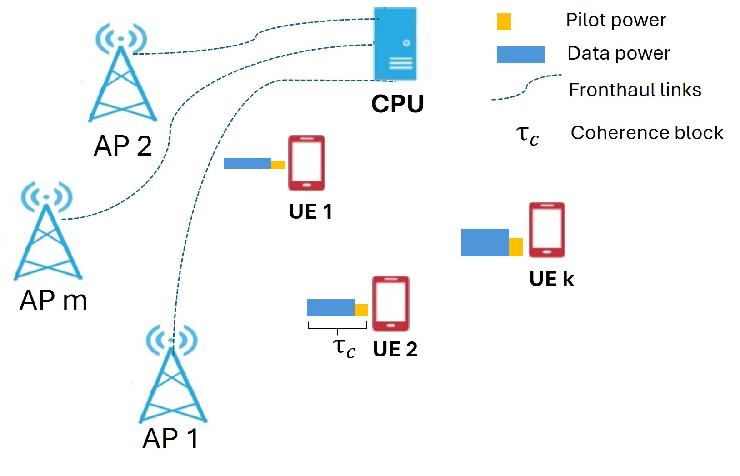}
	\caption{\add{Various widths and heights demonstrate different power for pilots and data for different users.}}
	\label{CF-mMIMO}
\end{figure} 
\section{System Model}

As shown in Fig.~\ref{CF-mMIMO}, we consider a CF-mMIMO system with $M$ randomly distributed APs that serve $K$ single-antenna users indexed by $ \mathcal{K} \in \{1,\cdots, K\}$, assuming that $K \ll M$. All APs are equipped with $N$ antennas and connected to a CPU via unconstrained fronthaul links. The system operates in time-division duplex mode, and all APs and users are perfectly synchronized. The channel $\mathbf{g}_{mk} \in \mathbb{C}^N$ between the $m^\text{th}$ AP and the $k^\text{th}$ user is modeled as
\begin{align}
    \mathbf{g}_{mk}= (\beta_{mk})^{1/2} \mathbf{h}_{mk}
\end{align}
where $\beta_{mk}$ and $\mathbf{h}_{mk} \in \mathcal{CN}(0,\mathbf{I}_N)$ represent the large-scale fading and small-scale fading coefficients, respectively. We assume that at each of the $m^\text{th}$ AP, the local statistical channel state information (CSI) of all connected links is perfectly known, as it can be estimated using standard methods \cite{sanguinetti2013random}. We consider a system that operates in the block-fading channel model, where the time and frequency plane is divided into coherence blocks, $\tau_c$, in which $\tau_p$ and $\tau_c-\tau_p$ dedicate the uplink pilot, and data transmission symbols, respectively.

\subsection{Uplink Training and Channel Estimation}
In this paper, we consider a CF-mMIMO network where a subset of APs in the network serves each user. We assume that each UE selects its pilot sequence randomly, which can be conducted in a fully distributed manner. Since the number of symbols for the training is $\tau_p$, only $\tau_p$  orthogonal pilot sequences are available in the network, so for $K> \tau_p$, multiple UEs have to share the same pilot sequence. \add{This random pilot assignment approach is straightforward but often results in nearby users sharing the same pilot sequence. Consequently, these users experience degraded performance due to substantial pilot contamination effects. To address pilot contamination, recent studies have proposed various algorithms to minimize its impact \cite{gottsch2021impact,gottsch2022subspace, polegre2021pilot,osawa2023overloaded}. These algorithms are designed to enhance the quality of communication by reducing interference caused by pilot reuse, particularly in dense multi-user environments. However, the focus of this paper is not based on the pilot decontamination algorithm, and we assume significant pilot contamination in this study due to random pilot assignment, allowing us to evaluate the proposed method under worst-case conditions. }

We consider $\tau_p$ mutually orthogonal pilot sequences as $\sqrt{\tau_p} \boldsymbol{\phi} \in \mathbb{C}^{\tau_p \times 1}$ where $\| \boldsymbol{\phi} \| ^2=1$ and the channel estimation is performed at the APs. We denote $\mathcal{A}_k$ the set of APs that serve the $k^\text{th}$ user. Let $\mathcal{S}_t \subset \{1,\cdots, K \} $ denote the subset of UEs assigned to pilot $t = \{1,\cdots, \tau_p \}$. By assigning the pilot sequence $\sqrt{\tau_p} \boldsymbol{\phi}_k$ for the $k^\text{th}$ user, the received pilot signal matrix at the $m^\text{th}$ AP  can be expressed as follows
\begin{equation} \label{RPS}
\mathbf {Y}_{mt}^{\mathrm {p}} = \sum _{k \in \mathcal{S}_t} (\tau_p p_{k}^{\mathrm {p}})^{1/2} \mathbf {g}_{mk} { \boldsymbol {\phi }}^{\mathrm {H}}_{k} + \mathbf {W}_{mt}^{\mathrm {p}}, 
\end{equation}
where $p_k^\mathrm{p}$ is the pilot power for user $k$ and $\mathbf {W}_{mt}^{\mathrm {p}} \in \mathbb{C}^{N \times \tau_p}$ represent the complex white Gaussian noise matrix, which has independent and identically distributed elements and is circularly symmetric $\mathcal{CN}(0, \sigma^2 \mathbf{I}_N)$, where $\mathbf{I}_N$ is $N \times N $ the identity matrix. To remove the interference from UEs, $\mathbf {Y}_{mt}^{\mathrm{p}}$ is projected onto $\boldsymbol{\phi}_k^\mathrm{H}$ which yields  $\mathbf{y}_{mt}^\mathrm{p}=\mathbf {Y}_{mt}^{\mathrm{p}} \boldsymbol{\phi}_k$, and the minimum mean-square error (MMSE) is utilized to estimate the channel as \cite{bjornson2017massive}
\begin{align*} 
\hat { \mathbf {{g}}}_{mk} = \dfrac{ \mathbb{E} \{ \mathbf{y}_{mk}^\mathrm{p} \mathbf {g}_{mk}^* \} }{ \mathbb{E} \{ \mathbf{y}_{mk}^\mathrm{p} (\mathbf{y}_{mk}^\mathrm{p})^\mathrm{H} \}} \mathbf{y}_{mt}^\mathrm{p} = (\tau_p p_{k}^{\mathrm{p}})^{1/2}\beta _{mk} {\theta}_{mk}^{-1} \mathbf{y}_{mt}^\mathrm{p}, 
\end{align*}
where $\hat { \mathbf {{g}}}_{mk}$ is the channel estimation and $\theta_{mk}$ is defined as
\begin{equation}  
   \theta_{mk} = \mathbb{E} \{ \mathbf{y}_{mk}^\mathrm{p} (\mathbf{y}_{mk}^\mathrm{p})^\mathrm{H} \} =  \tau_p \sum \limits _{j \in \mathcal{S}_t} p_{j}^{\mathrm {p}}\beta _{mj}| { \boldsymbol {\phi }}^{\mathrm {H}}_{k} \boldsymbol {\phi }_{j}|^{2} + \sigma^2.
\end{equation}
\indent The interference caused by the pilot-sharing UEs in $\mathcal{S}_t$ as indicated in \eqref{RPS} leads to pilot contamination, which degrades the system performance. Therefore, pilot contamination degrades the estimation performance, making coherent transmission less effective, and the estimates $\hat{\mathbf{g}}_{mk}$ for $k \in \mathcal{S}_t$ become correlated, leading to additional interference \cite{bjornson2017massive}. 

\subsection{Uplink Data Transmission and Achievable SE}
During the uplink data transmission, all users send their data to the APs using the same time-frequency resource. The received signal $\mathbf {y}_m^{\mathrm {d}} \in \mathbb{C}^N$, at the $m^\text{th}$ AP is given by
\begin{equation} 
\mathbf {y}_m^{\mathrm {d}} = \sum _{k=1}^{K} ({p_{k}^\mathrm {d}})^{1/2} \mathbf {{g}}_{mk}q_{k} + \mathbf {w}_{m}^{\mathrm {d}}, 
\end{equation}
where ${p_{k}^\mathrm {d}}$ is the transmit data power, $q_k$ with $\mathbb{E} \{ | q_k |^2 | \}=1$, is the data symbol transmitted by $k^\text{th}$ user, and $\mathbf {{w}}_{m}^{\mathrm {d}} \in \mathcal{CN}(0,\sigma^2 \mathbf{I}_N)$ is the additive white Gaussian noise at the receiver. For signal detection in the first stage, the $m^\text{th}$ AP utilizes the maximum ratio combining (MRC) technique in which the received signal, $\mathbf {y}_m^{\mathrm {d}}$ is multiplied by the Hermitian transpose of channel estimate,  $\hat { \mathbf {{g}}}_{mk}^\mathrm{H}$ and transmits $\hat { \mathbf {g}}_{mk}^\mathrm{H} \mathbf {y}_m^{\mathrm {d}} $ to the CPU through the backhaul link for all users. Then, for each user, $k$, the received products are combined as
\begin{align} \label{Recieved vector}
    y_k = \sum_{m=1}^M \hat { \mathbf {g}}_{mk}^\mathrm{H}& \mathbf {y}_m^{\mathrm {d}} 
     = ({p_{k}^\mathrm {d}})^{1/2} \big(\sum_{m=1}^M \hat { \mathbf {{g}}}_{mk}^\mathrm{H} \mathbf {g}_{mk} \big) q_{k} \nonumber \\ &+ \sum _{\substack{j=1\\ j\ne k}}^{K} ({p_{j}^\mathrm {d}})^{1/2} \big(\sum_{m=1}^M  \hat { \mathbf {g}}_{mk}^\mathrm{H} \mathbf {g}_{mj}\big) q_{j} + w_k^{\prime},
\end{align}
where $w_k^{\prime} = \sum_{m=1}^M \hat { \mathbf {g}}_{mk}^\mathrm{H} \mathbf {w}_{m}^{\mathrm {d}} $ represents the noise. In the second stage, the CPU gathers the local data estimates from all APs and combines them to create a final estimate of the UE data. The CPU calculates its estimate using a linear combination of the local estimates as
\begin{align} \label{Recieved vector estimate}
    \hat{y}_k &= \sum_{m=1}^M a_{mk}^* \hat { \mathbf {g}}_{mk}^\mathrm{H}  \mathbf {y}_m^{\mathrm {d}} \nonumber \\
     &= ({p_{k}^\mathrm {d}})^{1/2} \big(\sum_{m=1}^M a_{mk}^* \hat { \mathbf {{g}}}_{mk}^\mathrm{H} \mathbf {g}_{mk} \big) q_{k} \nonumber \\ &+ \sum _{\substack{j=1\\ j\ne k}}^{K} ({p_{j}^\mathrm {d}})^{1/2} \big(\sum_{m=1}^M a_{mk}^* \hat { \mathbf {g}}_{mk}^\mathrm{H} \mathbf {g}_{mj}\big) q_{j} + w_k^{\prime\prime},
\end{align}
where $w_k^{\prime\prime} = \sum_{m=1}^M a_{mk}^* \hat{\mathbf {g}}_{mk}^\mathrm{H} \mathbf {w}_{m}^{\mathrm {d}} $ and $a_{mk} \in \mathbb{C}$ defines the weight assigned by the CPU to the local signal estimate that the $m^\text{th}$ AP has of the signal from the $k^\text{th}$ user.
In this paper, we have assumed that the large-scale fading weight vector of user $k$ for all APs is deterministic.\\ 
\indent For simplicity, we define  the vector $\mathbf{u}_{jk} \in \mathbb{C}^M$ formed as
\begin{align} \label{New Received vector}
    \mathbf{u}_{jk} = [ \hat { \mathbf {g}}_{1k}^\mathrm{H} \mathbf {g}_{1j}, \cdots, \hat { \mathbf {g}}_{Mk}^\mathrm{H} \mathbf {g}_{Mj} ]^T,
\end{align}
where it represents the vector containing the combined receive channels between $j^\text{th}$ user and all APs serving $k^\text{th}$ user. Using \eqref{New Received vector}, the received signal at $k^\text{th}$ user in \eqref{Recieved vector estimate} can be expressed as 
\begin{align} \label{hat Recieved vector estimate}
    \hat{y}_k = ({p_{k}^\mathrm {d}})^{1/2} \mathbf{a}_{k}^\mathrm{T} \mathbf{u}_{kk}  q_k + \sum _{\substack{j=1\\ j\ne k}}^{K}  ({p_{j}^\mathrm {d}})^{1/2} \mathbf{a}_{k}^\mathrm{T} \mathbf{u}_{jk} q_j + w_k^{\prime\prime}.
\end{align}
\add{where $\mathbf{a}_k = [a_{1k}, \cdots, a_{mk}, \cdots, a_{Mk}]^\mathrm{T} $.} It can be observed from  \eqref{hat Recieved vector estimate} that the effective channels of the different users are represented by $(\mathbf{a}_k \mathbf{u}_{jk}: j=1, \cdots, K)$, as it has the structure of a single antenna channel. In order to compute the SE, we assume that the average of the effective channel, $\mathbb{E} \{ \mathbf{a}_k \mathbf{u}_{kk} \} $ is non-zero and deterministic, even though the effective channel $\mathbf{a}_k \mathbf{u}_{kk}$ is unknown at the CPU \cite{demir2021foundations}. Then, it can be assumed to be known. Therefore, the achievable uplink SE of the $k^\text{th}$ user can be written as follows
\begin{equation} \label{Original Rate}
\text{SE}_{k} = (1 - \frac{\tau_p}{\tau_{\text {c}}})\log _{2}\left ({1 +  \mathrm{SINR}_{k}}\right),
\end{equation}
where the SINR of the $k^\text{th}$ user is presented by \cite{bjornson2020scalable} 
\begin{align} \label{SINR Def}
 &\mathrm{SINR}_k \nonumber \\ & =\frac{{p_{k}^\mathrm {d}}\left| \mathbf{a}_{k} \mathbb{E}  \{\mathbf{u}_{kk}\}\right|^2}{ \mathbf{a}_{k}^\mathrm{T} \big(\sum_{\substack{j=1}}^K {p_{j}^\mathrm {d}} \mathbb{E}\{\mathbf{u}_{jk}\mathbf{u}_{jk}^\mathrm{H}\} - {p_{k}^\mathrm {d}} \mathbb{E}\{\mathbf{u}_{kk} \} \mathbb{E}\{\mathbf{u}_{kk}^\mathrm{H}\} + \mathbf{F}_k \big) \mathbf{a}_{k}},
\end{align}
where $\mathbf{a}_{k} = [1, \cdots, 1]^\mathrm{T} \in \mathbb{R}^M$ and
\begin{equation} \label{ukk}
    [\mathbb{E} \{\mathbf{u}_{kk}\}]_m = N \tau_p p_k^\mathrm{p}  \beta_{mk}^2 {\theta}_{mk}^{-1},
\end{equation}
\begin{align} \label{Fk}
    \mathbf{F}_k &= \sigma^2 \text{diag}\big( \mathbb{E} \lbrace \| \mathbf{g}_{1k} \|^2 \}, \cdots, \mathbb{E} \lbrace \| \mathbf{g}_{Mk} \|^2 \} \big) \in \mathbb{R}^{M \times M}
\end{align}
\begin{align} \label{ujk}
    [\mathbb{E}  \{\mathbf{u}_{jk} & \mathbf{u}_{jk}^\mathrm{H}\}]_{mm} = 
 \nonumber \\ &  N \tau_p p_k^\mathrm{p}  \beta_{mj} \beta_{mk}^2 \boldsymbol{}{\theta}_{mk}^{-1} 
  + N^2 p_k^\mathrm{p} p_j^\mathrm{p} \big( \tau_p \beta_{mj}  {\theta}_{mk}^{-1} \beta_{mk} \big)^2.
\end{align}
\textit{Proof:} Please refer to Appendix A \eqref{appendix A}. \\
By substituting these results into \eqref{SINR Def} and performing the necessary mathematical manipulations, we can reformulate the SINR as follows:
\begin{align} \label{SINR closed form}
&\mathrm{SINR}_k \nonumber \\& = \dfrac{N \tau_p p_k^\mathrm{d} p_k^\mathrm{p}  \sum_{m=1}^M\beta_{mk}^2 {\theta}_{mk}^{-1} }{ \sum_{m=1}^M \big( N \tau_p \sum _{\substack{j=1\\ j\ne k}}^{K} p_j^\mathrm{d} p_j^\mathrm{p} \beta_{mj}^2 {\theta}_{mk}^{-1} +  \sum_{j=1}^K p_j^\mathrm{d}  \beta_{mj}  + \sigma^2\big)}.
\end{align}

\begin{figure}[!]
	\centering
	\includegraphics[width=2.7in,height=0.9in]{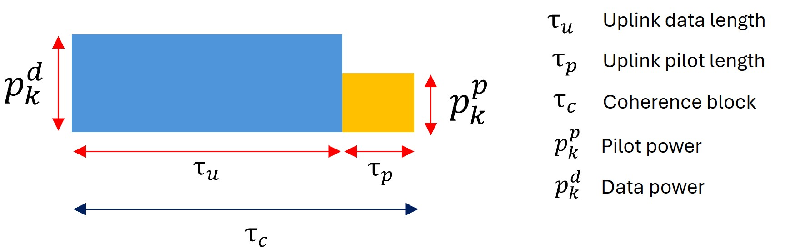}
	\caption{Demonstration of pilot and data power allocation based on $\tau_p$ and $\tau_u = \tau_c-\tau_p$ for uplink}
	\label{Pilot Data}
\end{figure} 

\add{\subsection{Pilot and data power constraint}
In CF-mMIMO, determining the per-user maximum transmit energy depends on system requirements like spectral efficiency, coherence block length, and interference management. Generally, the maximum transmit budget per user (often denoted as \(P_\text{max}\)) is determined based on the hardware’s power budget, regulatory constraints, and the need to mitigate interference. Studies suggest optimal power control strategies, including fractional or adaptive transmit power adjustments, help maintain performance within the system’s maximum energy limits, balancing data and pilot power for reduced interference and improved coverage. Thus, we can write
\begin{equation} \label{tau_c constraint}
   \tau_p p_k^\mathrm{p} + (\tau_c - \tau_p)p_k^\mathrm{d} \le E_\text{max} \Rightarrow \dfrac{\tau_p p_k^\mathrm{p} + (\tau_c - \tau_p)p_k^\mathrm{d}}{\tau_c} \le P_\text{max}.
\end{equation}
where $E_\text{max} = \tau_c P_\text{max}$. This inequality implies that pilot power (\(p_k^\mathrm{p}\)) and data power (\(p_k^\mathrm{d}\)) are jointly constrained to ensure the total average transmit power remains within \(P_{\text{max}}\). This relationship allows the system to manage the balance between pilot training and data transmission within the coherence block, which is important for improving spectral efficiency and mitigating interference. Fig.~\ref{Pilot Data} illustrates the allocation of pilot and data power based on the pilot sequence $\tau_p$ and the coherence block duration, $\tau_c$. This figure serves as a heuristic example aimed at providing an intuitive understanding of the scenario under consideration rather than offering a rigorous analysis. \\
\indent On the other hand, the coherence block duration, $\tau_c$ is influential in determining how transmit power is allocated in communication systems, particularly in multi-user scenarios. Although a UE’s transmit power is constrained by its hardware, such as amplifier efficiency and power limits, the coherence block length indirectly affects how efficiently this power can be used. Longer coherence blocks allow for more data symbols per channel estimation, enhancing spectral efficiency by reducing the time spent on training. In contrast, shorter coherence blocks, caused by high mobility or challenging environments, require more frequent re-estimation, limiting the potential efficiency for a given power budget. Additionally, shorter coherence durations make managing inter-user interference more complex, often requiring higher power levels to maintain service quality. Therefore, the duration of the coherence block can impact the power control strategies, ultimately influencing overall system performance. Moreover, the significance of $\tau_p$ is to quantify how much energy each terminal spends on pilots in each coherence interval \cite{marzetta2016fundamentals}. \\
\indent Based on this information, our objective in the following section is to utilize two distinct constraints—one for pilot power and the other for data power:
\begin{align}
    &p_k^\mathrm{p}  \leq \dfrac{\tau_c P_{\text{max}} - (\tau_c-\tau_p)p_k^\mathrm{d}}{\tau_p}, \\
    &p_k^\mathrm{d} \leq \dfrac{\tau_c P_{\text{max}} - \tau_p {p}_k^\mathrm{p}}{(\tau_c-\tau_p)}.    
\end{align}
These constraints are designed to ensure efficient resource allocation while maintaining system performance. The pilot power constraint will address the proper power allocation for channel estimation, minimizing interference and ensuring accurate channel state information. Meanwhile, the data power constraint will optimize the power used for data transmission to enhance spectral efficiency and mitigate interference. Together, these constraints form the foundation for a robust power control strategy tailored to the system’s requirements.}

\section{Pilot and data power Control}
In this section, we introduce the pilot and data PC technique. It determines the transmission power allocated by each user and the proportion of this power dedicated to pilot signals and data transmission. To this end, we introduce a novel constraint based on the maximum transmit budget power for the user and the coherence block and develop an iterative algorithm to minimize the NMSE. Subsequently, we employ a scheme based on McCormick relaxation to adjust the pilot power according to real-time channel conditions. Following this, we use GP to solve the max-min fairness optimization problem and allocate data power for each user. The algorithm aims to maximize the system's performance, including spectral efficiency and fairness, while adhering to specific constraints, such as total transmit power. Notably, this methodology differs from others by iteratively updating and adjusting both power levels for each channel realization.

\subsection{Pilot Power control}

During the training phase, the AP estimates the channels using the user pilots. However, due to non-orthogonality among pilot sequences, a pilot signal from one user can adversely affect the channel estimate of other users. This is known as the pilot contamination effect and can be particularly severe in CF-mMIMO systems that aim to serve multiple users simultaneously with the same time-frequency resource. Therefore, it is crucial to mitigate pilot contamination. To this end, we introduce iterative pilot power coefficients that can improve channel estimation accuracy during the training phase.\\
\indent It is known that the average channel gain influences the MSE value. Although a stronger channel may exhibit larger absolute errors than a weaker channel, the relative size of the error is more critical than its absolute magnitude. Consequently, estimation accuracy is evaluated using the NMSE. Specifically, for the channel between the $m^\text{th}$ AP and the $k^\text{th}$ UE, the NMSE is defined as:
\begin{align} \label{NMSE}
    \text{NMSE}_{mk} &= \dfrac{\mathbb{E} \{ \| \mathbf{g}_{mk} - \hat{\mathbf{g}}_{mk} \|^2 \}}{\mathbb{E} \{ \| \mathbf{g}_{mk} \|^2 \}} = 1 - \dfrac{N p_k^\mathrm{p} \tau_p \beta^2_{mk} \theta_{mk}^{-1}}{N \beta_{mk}} \nonumber \\
    & = 1 - \dfrac{ \tau_p p_k^\mathrm{p} \beta_{mk}}{   \tau_p \sum _{\substack{j=1}}^{K} p_{j}^{\mathrm {p}}\beta _{mj}| { \boldsymbol {\phi }}^{\mathrm {H}}_{k} \boldsymbol {\phi }_{j}|^{2} + \sigma^2}.
\end{align}
\indent The system calculates the variance of the channel estimation error between the $m^\text{th}$ AP and the $k^\text{th}$ user and presents it as a relative value. It can be seen from \eqref{NMSE} that NMSE reduces as the channel estimation error ($  \mathbf{g}_{mk} - \hat{\mathbf{g}}_{mk}$) of the $k^\text{th}$ user decreases. The proposed algorithm aims to allocate pilot power for each user and provide uniformly good service to all network users. We formulate a min-max optimization problem designed to minimize the maximum user-NMSE based on a novel constraint based on the maximum transmit budget at the users and the coherence time to achieve this objective as follows
\begin{subequations} \label{Pilot original problem}
 \begin{align} 
\textbf{P1}: &\min _{\lbrace p_k^\mathrm{p}\rbrace } \, \max _{k=1, \ldots, K} \sum _{m \in \mathcal{A}_k} \text{NMSE}_{mk} \nonumber  \\ &\text{subject to}  \quad   p_k^\mathrm{p}  \leq \dfrac{\tau_c P_{\text{max}} - (\tau_c-\tau_p)p_k^\mathrm{d}}{\tau_p}  \label{Pilo2 b} \\
& \qquad  \qquad  \quad \epsilon \leq p_k^\mathrm{p}, \quad \forall k\in \mathcal{K}
\end{align}
\end{subequations}
where $P_\text{max}$ denotes the maximum transmit budget available for each user. Constraint \eqref{Pilo2 b} maintains a balance between the power assigned for pilot and data transmission, allowing the total power usage to stay within the prescribed limits. The inequality reflects a trade-off between utilizing power for pilot transmission to estimate channel parameters and allocating power for data transmission for information transfer.  \\
To solve the optimization problem in \eqref{Pilot original problem}. without loss of generality, problem \textbf{P1} can be rewritten by introducing a new slack variable as 
\begin{subequations} \label{Pilot problem}
\begin{align} 
\textbf{P2}: &\min _{\lbrace p_k^\mathrm{p}, \nu \rbrace } \, \nu  \nonumber \\ &\text{subject to} \nonumber \\
&\sum_{m \in \mathcal{A}_k} \Big(1 - \dfrac{ \tau_p p_k^\mathrm{p} \beta_{mk}}{ \tau_p \sum _{\substack{j=1}}^{K} p_{j}^{\mathrm {p}}\beta _{mj} \left| { \boldsymbol {\phi }}^{\mathrm {H}}_{k} \boldsymbol {\phi }_{j} \right|^{2} + \sigma^2} \Big) \leq \nu \label{nonconvexity} \\
\quad  & \epsilon \leq p_k^\mathrm{p}  \leq \dfrac{\tau_c P_{\text{max}} - (\tau_c-\tau_p)p_k^\mathrm{d}}{\tau_p}, \quad \forall k\in \mathcal{K} \label{new constraint}   
\end{align}
\end{subequations}
However, the constraint involving the sum of ratios in \eqref{nonconvexity} is non-convex due to the division by a quadratic term. Thus, problem \textbf{P2} is non-convex. To tackle this non-convexity, we present a computationally efficient algorithm to convert the concave constraint into a convex one based on McCormick relaxation. This technique introduces auxiliary variables and linear constraints to approximate the concave term with a piecewise linear function. The McCormick relaxation algorithm is utilized to address non-convex optimization problems by creating convex and concave relaxations to approximate the non-convex feasible region. We rewrite \eqref{nonconvexity} as
\begin{align}
& 1-\frac{\tau_p p_k^{\mathrm{p}} \beta_{m k}}{\tau_p \sum_{j=1}^K p_j^{\mathrm{p}} \beta_{m j}\left|\boldsymbol{\phi}_k^H \boldsymbol{\phi}_j\right|^2+\sigma^2}=q_{mk}. \quad \forall m, k 
\end{align}
To linearize this term using McCormick relaxation, we can introduce auxiliary variables $z_{mk}$ and $v_{mk}$ to approximate $q_{mk}$:
\begin{align}
    z_{mk} &= \tau_p p_k^\mathrm{p} \beta_{mk} , \\
    v_{mk} &=  \tau_p \sum _{\substack{j=1}}^{K} p_{j}^{\mathrm {p}}\beta _{mj}| { \boldsymbol {\phi }}^{\mathrm {H}}_{k} \boldsymbol {\phi }_{j}|^{2} + \sigma^2, 
\end{align}
where $z_{mk} \ge 0$ and $v_{mk} \ge 0$. We then add linear constraints to ensure that $z_{mk}$ and $v_{mk}$  correctly approximate $q_{mk}$, and we reformulate the optimization problem as follows:
\begin{align} \label{P3}
\textbf{P3}: & \min _{\lbrace p_k^\mathrm{p}, \nu \rbrace } \, \nu \nonumber \\
& \text{subject to} \quad 
\begin{aligned}[t]
& \sum _{m \in \mathcal{A}_k} q_{mk} \leq \nu \\
& q_{mk} \le 1- \frac{z_{mk}}{v_{mk}} + (1-z_{mk}) \\
& q_{mk} \ge 1- \frac{z_{mk}}{v_{mk}} + (1-v_{mk}) \\
& \epsilon \leq p_k^\mathrm{p} \leq \dfrac{\tau_c P_{\text{max}} - (\tau_c-\tau_p)p_k^\mathrm{d}}{\tau_p}, \quad \forall k\in \mathcal{K}
\end{aligned}
\end{align}
This method allows for the creation of relaxations that can be solved efficiently and provide bounds for the optimal solution. These constraints and the objective function represent the linearized form of the original non-convex optimization problem \textbf{P2}. Thus, problem \textbf{P1} can be solved by solving a sequence of convex problem \textbf{P3}. Using yet another linear matrix inequalities parser (YALMIP) \cite{boyd2004convex} of Matlab toolbox, the optimized result is represented by the pilot power vector $\hat{\mathbf{p}}^{\mathrm{p}}_k $ which stacks all pilot power coefficients $\hat{p}_k^\mathrm{p}$.
\subsection{Data Power Control}

One important aspect of CF-mMIMO systems is their capability to provide consistently equal service to all users. We focus on the optimization problem of max-min fairness, which entails using the pilot power coefficients calculated in problem \textbf{P3}, ($\hat{\mathbf{p}}^{\mathrm{p}}_k$) and optimizing the data PC coefficient to maximize the minimum user rates. We denote the data power coefficient vector as $\mathbf{p}_k^\mathrm{d} = [p_1^\mathrm{d}, p_2^\mathrm{d}, \cdots, p_K^\mathrm{d}]$, then rewrite the uplink SE of $k^\text{th}$ user as
\begin{equation} \label{Original Rate}
\mathrm{SE}_k (\hat{\mathbf{p}}_k^\mathrm{p},\mathbf{p}_k^\mathrm{d}) = (1 - \frac{\tau_p}{\tau_c} )\log _{2}\left ( {1 +  \mathrm{SINR}_{k} (\hat{\mathbf{p}}_k^\mathrm{p},\mathbf{p}_k^\mathrm{d})} \right),
\end{equation}
and optimization problem is formulated as:
\begin{align} \label{original data}
\textbf{P4}: &\max _{\lbrace p_k^\mathrm{d}\rbrace } \, \min _{k=1, \ldots, K} \mathrm{SE}_{k} (\hat{\mathbf{p}}_k^\mathrm{p},\mathbf{p}_k^\mathrm{d}) \nonumber \\ 
&\text{subject to} \quad 
\begin{aligned}[t]
& 0 \leq p_k^\mathrm{d} \leq \dfrac{\tau_c P_{\text{max}} - \tau_p \hat{p}_k^\mathrm{p}}{(\tau_c-\tau_p)}, \quad \forall k \in \mathcal{K} 
\end{aligned}
\end{align}
We exploit GP (convex problem) to develop an efficient solution for Problem \textbf{P4}, which is defined in \eqref{original data}. Problem \textbf{P4} cannot be directly solved through the optimization software. By utilizing the slack variables, the optimization problem can be reformulated as 
\begin{subequations}\label{GP}
\begin{align}
& \textbf{P5} && \max _{{\lbrace p_k^\mathrm{d}\rbrace },t} \, t  \nonumber \\
& && \text{subject to}  &&t \leq \mathrm{SINR}_{k} (\hat{\mathbf{p}}_k^\mathrm{p},\mathbf{p}_k^\mathrm{d}),\\
& &&   &&0 \leq p_k^\mathrm{d}  \leq \dfrac{\tau_c P_{\text{max}} - \tau_p \hat{p}_k^\mathrm{p}}{(\tau_c-\tau_p)},   \quad \forall k \in \mathcal{K} \label{zeroone}, 
\end{align}
\end{subequations}
\textbf{Lemma 1}. \textit{Problem 5 (\textbf{P5}) can be solved using a GP, allowing for a globally optimal solution in polynomial time.}\\
\indent \textit{Proof:} Problem \eqref{GP} can be written as a GP in standard form \cite{boyd2004convex}. In \eqref{GP}, the objective function is already in the form of \textit{t}, a monomial and thus a valid posynomial\footnote{A function $q\left(y_1, \ldots, y_{T_1}\right)=\sum_{t=1}^{T_2} a_t \prod_{m=1}^{T_1} y_m^{b_{t, m}}$ is posynomial with $T_2$ terms $\left(T_2 \geq 2\right.$ ) if the coefficients $b_{t, m}$ are real numbers and the coefficients $a_t$ are nonnegative real numbers. When $T_2=1, q\left(y_1, \ldots, y_{T_1}\right)$ is a monomial.}. \add{The power constraints are expressed as monomials, while the SINR expressions can be rearranged as posynomial constraints as follows in Appendix B:
\begin{align}
N \tau_p p_k^\mathrm{d} p_k^\mathrm{p}& \sum_{m=1}^M \beta_{mk}^2 \theta_{mk}^{-1} \geq \\ \nonumber &t \sum_{m=1}^M \Big( N \tau_p \sum_{\substack{j=1 \\ j \ne k}}^{K} p_j^\mathrm{d} p_j^\mathrm{p} \beta_{mj}^2 \theta_{mk}^{-1} + \sum_{j=1}^K p_j^\mathrm{d} \beta_{mj} + \sigma^2 \Big).
\end{align}
Additionally, equation \eqref{zeroone} is in the form of a posynomial. Since geometric programs have a convex structure, \textbf{P5} can be efficiently solved in a centralized manner using the interior-point toolbox CVX \cite{boyd2004convex,grant2008cvx}.
This process must be repeated until the maximum number of iterations is reached.} \\
\indent The transmitted data power allocated to user $k$ is intricately influenced by several key factors. Firstly, the maximum transmit power, denoted by \( P_{\text{max}} \), sets a fundamental constraint on the power level that can be allocated to each user. Additionally, the coherence block duration, represented by \( \tau_c \), is pivotal in determining the permissible power allocation. This duration governs the temporal coherence of the channel, thereby influencing the power allocation strategy. Moreover, the estimated pilot power, denoted by $\hat{p}_k^\mathrm{p} $, is a crucial parameter calculated based on the minimization of the normalized mean square error for the channel. This estimation process ensures an optimal trade-off between pilot power and channel estimation accuracy, directly impacting the effectiveness of the power allocation scheme. Collectively, these factors underscore the intricate interplay between system parameters and optimization objectives in determining the transmitted data power for each user.

The key distinction between the proposed method and existing algorithms lies in their approach to updating data power based on channel states. Unlike the existing literature, where data power adjustments do not occur in tandem with each pilot PC cycle, our approach allows for observing the interdependent effects of updating data power on pilot power and vice versa. Previous literature typically separates the calculation of pilot power from the data power. Specifically, existing approaches often compute pilot power independently and subsequently utilize it to determine data power. In contrast, our proposed method directly incorporates CSI into the data power calculation. By considering the variance of the channel in which the pilot power is utilized, our approach offers a more integrated and adaptive strategy for power allocation, potentially leading to improved system performance and efficiency.

\subsection{Computational Complexity}

\add{To evaluate the computational requirements, we compute the complexity of each component for $K$ users and $|\mathcal{A}_k| \le M$ APs within the power control methods in terms of floating point operations (FLOPs) as follows. The computational complexity of the proposed iterative pilot power allocation (IPPA) method is explained in Algorithm 1, which solves the problem \textbf{P3} and the GP convex optimization problem, \textbf{P5} at each iteration. For pilot power calculation, we consider \textbf{P3}. The calculation process for $q_{mk}$ includes $z_{mk}$ with complexity of $\mathcal{O} (|\mathcal{A}_k| K)$ and $v_{mk}$ with complexity of $\mathcal{O} (2|\mathcal{A}_k| K \tau_p^2)$. At the same time, the division makes $\mathcal{O} (2|\mathcal{A}_k| K)$ and summation for each $k$ is $\mathcal{O} (|\mathcal{A}_k| K)$. On the other hand, constraints have totally $\mathcal{O} (6|\mathcal{A}_k| K^2)$. Thus, the totals complexity for solving \textbf{P3} is $ \mathcal{O} (2|\mathcal{A}_k|K\tau_p^2 +|\mathcal{A}_k|K + 2|\mathcal{A}_k|K + K|\mathcal{A}_k| + 6|\mathcal{A}_k|K^2) \approx \mathcal{O}(2|\mathcal{A}_k|K\tau_p^2 + 6|\mathcal{A}_k|K^2)$. For the data power optimization problem in \textbf{P5}, calculating SINR across all $K$ users yields $\mathcal{O}(4 |\mathcal{A}_k| K \tau_p^2 + 4| \mathcal{A}_k|K )$ and the iterative optimization process in GP solver incurs complexity of $\mathcal{O}(K^{3.5})$ while the complexity of setting up the constraints for GP becomes $\mathcal{O}(K^2)$. The total complexity over $T$ iterations for the proposed method becomes $\mathcal{O}(T \cdot \Big(6|\mathcal{A}_k|K\tau_p^2 + 6|\mathcal{A}_k|K^2 + K^{3.5} + K^2 +4| \mathcal{A}_k|K ) \Big) \approx \mathcal{O}(T \cdot K^{3.5})$.\\
In the no-pilot power allocation (NPPA) algorithm \cite{ngo2017cell}, only the data power is controlled while the pilot power is set to the maximum power under the pilot power constraint. The minimum SE is maximized as the optimization objective. 
This method, considering  $\mathcal{O}(|\mathcal{A}_k|K^2+|\mathcal{A}_k|K^2\tau_p)$ as the complexity for calculating SINR across all users, based on max-min power allocation with iterative bisection, has a total complexity of $\mathcal{O}(|\mathcal{A}_k|^2K + |\mathcal{A}_k|^2 + K |\mathcal{A}_k| \log |\mathcal{A}_k| + K|\mathcal{A}_k| + T |\mathcal{A}_k|K^2 + T|\mathcal{A}_k|K^2\tau_p)$ for $T = \log_2(\frac{t_\mathrm{max}-t_\mathrm{min}}{\kappa})$ iterations. Since this method is based on the bisection method, which results in a per-iteration complexity in the order of $\mathcal{O}(T \cdot K^4)$.\\
The controlled pilot power allocation (CPPA) algorithm \cite{mai2018pilot} is a scheme focused solely on controlling pilot power, while the uplink data power is set to its maximum within the data power constraint. It has an initial complexity of $\mathcal{O}(|\mathcal{A}_k| K^2)$ for computing the starting values based on algorithm 1 in \cite{mai2018pilot} and follows an inner iteration (over $T = \log2(\frac{t_\mathrm{max}-t_\mathrm{min}}{\kappa})$ iterations) with complexity $\mathcal{O}(|\mathcal{A}_k|K^2 + |\mathcal{A}_k|K^2 \tau_p)$ for each iteration. This process is repeated over $T_o$ outer iterations. Thus, the final complexity of the CPPA method is $\mathcal{O}(|\mathcal{A}_k|^2 K + |\mathcal{A}_k|^2 + K |\mathcal{A}_k| \log |\mathcal{A}_k| + K |\mathcal{A}_k| + |\mathcal{A}_k|K^2 + T_o T |\mathcal{A}_k|K^2 + |\mathcal{A}_k|K^2 \tau_p)$. Although the primary goal of this method is to minimize channel estimation error, it does not adjust the data power in response to changes in pilot power. The solution from the CPPA technique entails solving a convex optimization problem followed by applying the bisection method, resulting in a per-iteration complexity of $\mathcal{O}(T_o \cdot T \cdot K^4)$. }


\begin{algorithm}[t]
\caption{PC - UA Algorithm}
\label{alg_1}
\SetAlgoLined
\textbf{Input:} Set of APs $\mathcal{A}_k$, $ \mathcal{K} \in \{1,\cdots, k, \cdots, K\}$ users, large scale fading, $\beta_{mk}$  $\forall m \in \mathcal{A}_k$, $\forall k \in \mathcal{K}$ \;
 \textbf{Initialization:} $ \mathbf{p}^\mathrm{d}_k = [1, \cdots, 1] $, $\bf{\zeta} > 0 $\;
  \For{\text{channel realization}}{
       \Repeat{ end of iteration}{Solve Problem \textbf{P3} defined in \eqref{P3} based on the given constraints to obtain $\hat{\mathbf{p}}_k^\mathrm{p}$\;
       Set $ \hat{\mathbf{p}}_k^{\mathrm{P}_\text{new}} = \hat{\mathbf{p}}_k^\mathrm{p} $ \;
       Solve Problem \textbf{P5} defined in \eqref{GP} based on the given constraints to obtain $\hat{\mathbf{p}}_k^{\mathrm{d}_\text{new}}$ \; 
       \uIf{$\Vert \hat{\mathbf{p}}_k^{\mathrm{d}_\text{new}} - \hat{\mathbf{p}}_k^{\mathrm{d}_\text{old}} \Vert > \bf{\zeta} $}{
        $\hat{\mathbf{p}}_k^{\mathrm{d}^*} = \hat{\mathbf{p}}_k^{\mathrm{d}_\text{new}}$\;
       Break\;
        }
        \Else{$\hat{\mathbf{p}}_k^{\mathrm{d}_\text{old}} = \hat{\mathbf{p}}_k^{\mathrm{d}_\text{new}}$\;
        Go to Line 5\;
        }
          }
          \textbf{Output} $\hat{\mathbf{p}}_k^{\mathrm{d}^*}$, $\hat{\mathbf{p}}_k^{\mathrm{p}^*}$\;
 \For{\text{each user}}{
 \text{Utilize achieved powers} $\hat{\mathbf{p}}_k^{\mathrm{d}^*}$, $\hat{\mathbf{p}}_k^{\mathrm{p}^*}$ from line 16\;
  Calculate  $\text{SINR}_k$ from  \eqref{SINR closed form}\;
  Calculate   $\text{SE}_k$ from  \eqref{Original Rate}\;
  }
  \textbf{Output} $\text{SE}$
   }
\end{algorithm}


\section{Numerical Results}
This section presents numerical results to illustrate how the proposed approach enhances the scalability of CF-mMIMO systems (the $k^\text{th}$ user is served by only the APs in $\mathcal{A}_k$ ). We analyze a simulation scenario with $M$ APs, $K = 40$ UEs; each coherence block consists of $\tau_c=200$ samples, and the total number of distributed antenna elements is $MN$. We performed 500 channel realizations. It is assumed that APs and UEs are independently and uniformly distributed within an area of $1 \times 1$ $\text{km}^2$, and a wrap-around technique is used to prevent boundary effects and simulate a network with an infinite area. The large-scale fading coefficients are modeled by \cite{ngo2017cell}
\begin{equation*} \beta _{mk} = {\Gamma }_{mk} \cdot 10^{\frac{\sigma _\mathrm{sh} z_{mk}}{10}}, 
\end{equation*}
where $\Gamma_{mk}$ denotes the pathloss, and $10^{\frac{\sigma _\mathrm{sh} z_{mk}}{10}}$ represent the shadow fading with standard deviation $\sigma_\mathrm{sh} = 8$ dB, and $z_{mk} \sim \mathcal{N}(0,1) $. The path loss \(\Gamma_{mk}\) is modeled using a three-slope \cite{tang2001mobile}, which depends on the distance \(d_{mk}\) between the $m^\text{th}$ AP and the $k^\text{th}$ user. It is defined as follows:
\begin{itemize}
    \item For \(d_{mk} > d_1\): \begin{align}
  \Gamma_{mk} = -L - 35 \log_{10}(d_{mk}).
  \end{align}
  \item For \(d_0 < d_{mk} \leq d_1\):
  \begin{align}
  \Gamma_{mk} = -L - 15 \log_{10}(d_1) - 20 \log_{10}(d_{mk}).
  \end{align}
  \item  For \(d_{mk} \leq d_0\):
  \begin{align}
  \Gamma_{mk} = -L - 15 \log_{10}(d_1) - 20 \log_{10}(d_0).
  \end{align}
\end{itemize}
where \(d_0 = 10\)m and \(d_1 = 50\)m. Note that shadow fading is considered only when \(d_{mk} \geq d_1\), meaning there is no shadowing if \(d_{mk} < d_1\). $L$ is calculated as
\begin{align*} 
L &\triangleq 46.3 + 33.9 \log _{10} \left({f}\right) - 13.82 \log _{10} \left(h_{\text{AP}}\right)\\ & \qquad - \left(1.1 \log _{10} \left({f }\right) - 0.7\right)h_{k} + \left(1.56 \log _{10} \left({f }\right) - 0.8\right), 
\end{align*}
where $f = 1900$MHz is the carrier frequency,  $h_k=1.65$m and $h_\text{AP}=15$m are the height of user $k$ and AP $m$, respectively. The noise power is defined as
\begin{align}
    \text{noise power} = k_B \times T_0 \times B \times NF,
\end{align}
where $NF = 9$ dB is the noise figure, the bandwidth is fixed in $B= 20$ MHz, the noise temperature is $T_0=290$ (Kelvin), and $k_B = 1.381 \times 10^{-23}$ (Joule per Kelvin ) is the Boltzmann constant.

\subsection{Simulation results}
\begin{figure}[!]
\centering
\subfloat[$\tau_p = 5$]{\label{N100tau5} \includegraphics[height=2.2in]{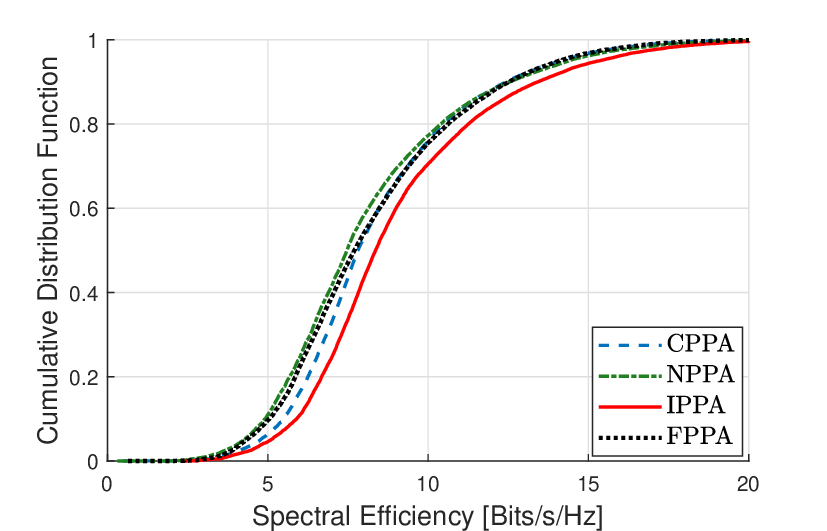}} 
\qquad
\subfloat[$\tau_p = 10 $]{\label{N100tau10} \includegraphics[height=2.2in]{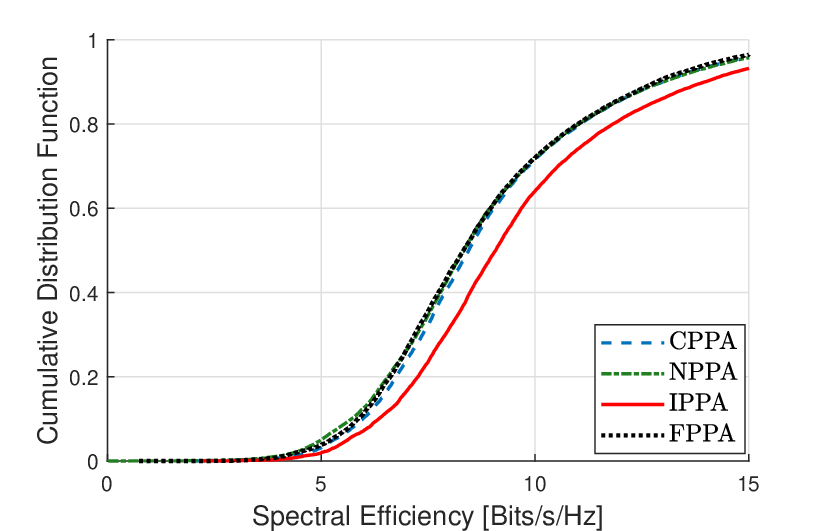}} 
\caption{CDF of the spectral efficiency for AP=100, N=1, K=40}
\label{N100}
\end{figure}

\indent We assess the effectiveness of the proposed iterative pilot power allocation (IPPA) scheme by comparing it with a no-pilot power allocation (NPPA) scheme \cite{ngo2017cell}, the controlled pilot power allocation (CPPA) scheme \cite{mai2018pilot}, and the fractional pilot power control (FPPA) scheme \cite{nikbakht2019uplink}. Note that the "NPPA" curves correspond to the cases where the users transmit with full power during the training phase. Moreover, in the proposed IPPC algorithm, we choose $\epsilon = 0.01$, and the maximum transmit budget  $P_\mathrm{max}$= 100 mW.
All considered schemes use the radio unit-antennas association scheme to ensure a fair comparison. This means that the distribution of APs decreases as the number of antennas increases ($MN=100$). We consider the max-min fairness power allocation scheme \textbf{P5} in \eqref{GP} for data transmission.\\
\indent Fig.~\ref{N100tau5} and Fig.~\ref{N100tau10} present the cumulative distribution function (CDF) of the SE per UE for the proposed IPPA versus the other tested algorithms when $\tau_p = 5$ and $\tau_p = 10$, respectively. In the FPPA, 60 percent of the UEs archive better SE, while 40 percent of the rest of the UEs in the network have close performance to the CPPA. In contrast, the proposed IPPA method consistently performs better than the other algorithms for both values of $\tau_p$. This highlights the effectiveness of the iterative pilot power allocation (IPPA) approach, especially in scenarios with a limited number of pilot sequences, demonstrating its clear advantages over other compared power allocation methods.\\
\indent Fig.~\ref{N100}, shows the cumulative distribution of the achievable rates for $M=25$ and $N=4$. Compared to the previous scenario, one can observe that the CPPA and FPPA algorithms have almost the same performance as the NPPA algorithm but fall short of the performance achieved by the proposed IPPA method. The proposed method achieves the best performance for values of $\tau$ because it minimizes the maximum mean square error in the proposed algorithm, improving the quality of channel estimation and mitigating pilot contamination. At the same time, data PC is based on maximizing the minimum SE, and pilot power is adaptively controlled to improve the SE further. 

\begin{figure}[!]
\centering
\subfloat[$\tau_p = 5$]{\label{WPCtau5} \includegraphics[height=2.25in]{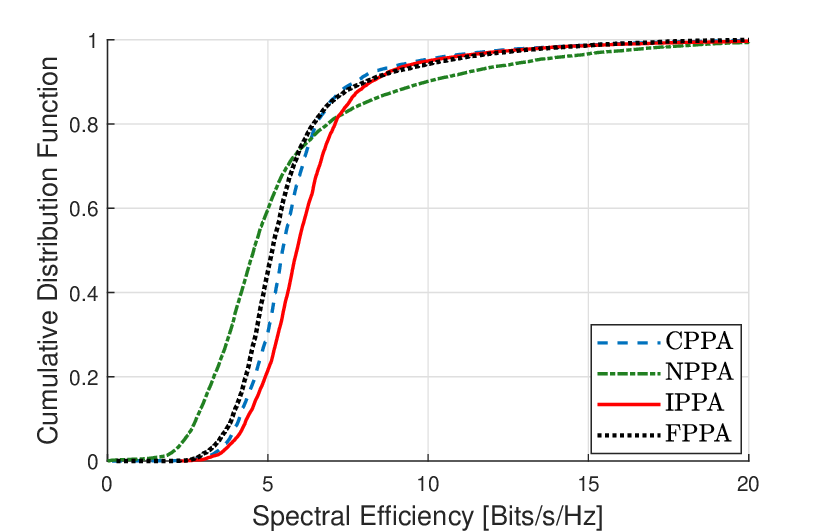}}
\qquad
\subfloat[$\tau_p = 10 $]{\label{WPCtau10} \includegraphics[height=2.25in]{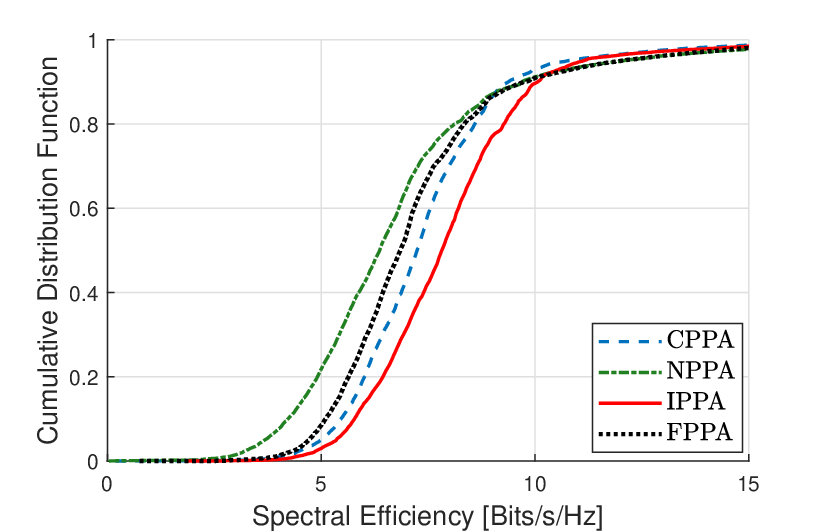}}
\caption{CDF of the spectral efficiency  with  AP=25, N=4, K=40}
\label{N25}
\end{figure}
\begin{figure}[!]
	\centering
	\includegraphics[width=3.8in,height=2.5in]{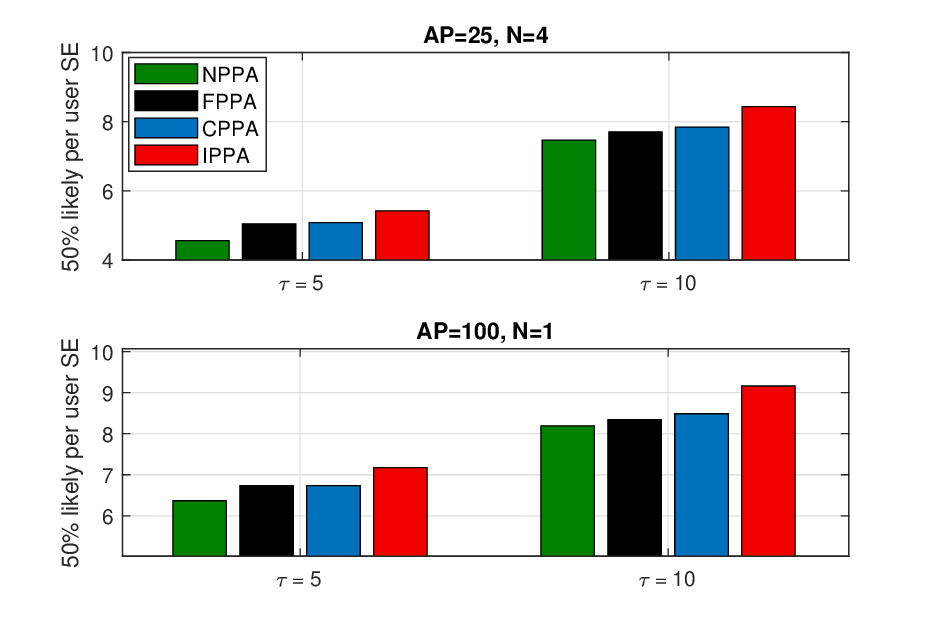}
	\caption{50\%  likely per user spectral efficiency}
	\label{50 likely}
\end{figure}

\indent Fig.~\ref{50 likely} illustrates the 50 percent likely per-user SE as a function of the number of APs for three schemes as the number of APs is changed. Each scheme is evaluated under two conditions for $\tau_p = 5$ and $\tau_p = 10$. The results show that SE increases with the number of APs for all schemes. Our proposed IPPA method consistently outperforms other compared methods, such as NPPA, CPPA, and FPPA, in the 50 percentile SE per user. This improvement is observed regardless of the number of APs or pilot symbols, demonstrating the robustness and scalability of the IPPA method across various network configurations.

\section{Conclusion}
In this work, we developed a simple yet effective algorithm that iteratively minimizes the NMSE while incorporating additional constraints for simultaneous PC. We presented an efficient iterative pilot and data PC scheme tailored explicitly for CF-mMIMO systems. This scheme maximizes spectral efficiency, ensures reliable communication links, and meets users' QoS requirements. The proposed scheme leverages advanced optimization techniques to allocate pilot and data powers based on channel conditions adaptively. Our novel approach distinguishes itself by updating and adjusting pilot and data powers in each channel realization. This methodology sets our work apart from existing methods, considering joint and updates of both power types. Through simulations, we demonstrated the effectiveness and efficiency of the proposed scheme in enhancing system performance and achieving optimal power allocation in CF-mMIMO deployments.


\section*{Appendix A} \label{appendix A}
Proof of \eqref{ukk} - \eqref{ujk}
\begin{align}
[\mathbb{E} \{\mathbf{u}_{jk}\}]_m &= \mathbb{E} \{ \mathbf{g}_{mk}^\mathrm{H} \mathbf{g}_{mj} \} \stackrel{(i)}{=} \mathbb{E} \{ \hat{\mathbf{g}}_{mj} \hat{\mathbf{g}}_{mk}^\mathrm{H} \}  \nonumber \\ &= \begin{cases}  \sqrt{p_k^\mathrm{p} p_j^\mathrm{p}} \tau_p ( \beta_{mj} \theta_{mk}^{-1} \beta_{mk}) & j \in \mathcal{I}_k \\ 
0  & j \notin \mathcal{I}_k
\end{cases}\tag{8}
\end{align}
where $\mathcal{I}_k$ is the set of UEs that use the same pilot as UE $k$. Since $\mathbf{g}_{mj} = \hat{\mathbf{g}}_{mj} +\Tilde{\mathbf{g}}_{mj}$ and  $\mathbf{g}_{mk} = \hat{\mathbf{g}}_{mk} +\Tilde{\mathbf{g}}_{mk}$, then $(\textit{i})$ follows the fact that   $\Tilde{\mathbf{g}}_{mj}$ (channel estimation error) and $\hat{\mathbf{g}}_{mk}$ are independent. The equality in \eqref{Fk} follows as
\begin{align}
    [\mathbf{F}_k]_{mm} &= \sigma^2 \mathbb{E} \{ \hat{\mathbf{g}}_{mk} \hat{\mathbf{g}}_{mk}^\mathrm{H} \}  = \sigma^2 [\mathbb{E} \{\mathbf{u}_{kk}\}]_m \nonumber \\
    & = \sigma^2 p_k^\mathrm{p}  \tau_p ( \beta_{mk} \theta_{mk}^{-1} \beta_{mk}).
\end{align}
To compute \eqref{ujk}, we observe that $ \mathbb{E} \{ [\mathbf{u}_{mj}]_m [\mathbf{u}_{mj}^*]_r \} =  \mathbb{E} \{ [\mathbf{u}_{mj}]_m \}  \mathbb{E} \{ [\mathbf{u}_{mj}^*]_r \}$ for $r \neq m$ since the the channels of different APs are independent. Hence, it only remains to compute
\begin{align}
     [\mathbb{E}  \{\mathbf{u}_{jk}  \mathbf{u}_{jk}^\mathrm{H}\}]_{mm} &= \mathbb{E}\left\{\hat{\mathbf{g}}_{mk}^{\mathrm{H}} \mathbf{g}_{mj} \mathbf{g}_{mj}^{\mathrm{H}} \hat{\mathbf{g}}_{mk} \right\} \nonumber \\ &= \mathbb{E}\left\{\mathbf{g}_{mj} \mathbf{g}_{mj}^{\mathrm{H}} \hat{\mathbf{g}}_{mk} \hat{\mathbf{g}}_{mk}^{\mathrm{H}}\right\}.
\end{align}
If $ j \notin \mathcal{I}_k$, we can utilize the independence of $\hat{\mathbf{g}}_{mk}$ and $\mathbf{g}_{mj}$ to obtain

\begin{align}
     \mathbb{E}\left\{\mathbf{g}_{mj} \mathbf{g}_{mj}^{\mathrm{H}} \hat{\mathbf{g}}_{mk} \hat{\mathbf{g}}_{mk}^{\mathrm{H}}\right\}  &= \mathbb{E}\left\{\mathbf{g}_{mj} \mathbf{g}_{mj}^{\mathrm{H}}\right\}  \mathbb{E}\left\{\hat{\mathbf{g}}_{mk} \hat{\mathbf{g}}_{mk}^{\mathrm{H}}\right\} \nonumber
     \\ &=  p_k^\mathrm{p} \tau_p \beta_{mj} \beta_{mk} \theta_{mk}^{-1} \beta_{mk}.
\end{align}
If $j \in \mathcal{I}_k $, we can write that
\begin{align}
    &\mathbb{E}\left\{\mathbf{g}_{mj} \mathbf{g}_{mj}^{\mathrm{H}} \hat{\mathbf{g}}_{mk} \hat{\mathbf{g}}_{mk}^{\mathrm{H}}\right\} \stackrel{(ii)}{=} \nonumber \\  &\mathbb{E} \{ \hat{\mathbf{g}}_{mj} \hat{\mathbf{g}}_{mj}^\mathrm{H} \hat{\mathbf{g}}_{mk} \hat{\mathbf{g}}_{mk}^\mathrm{H} \} +  \mathbb{E} \{ \tilde{\mathbf{g}}_{mj} \tilde{\mathbf{g}}_{mj}^\mathrm{H} \hat{\mathbf{g}}_{mk} \hat{\mathbf{g}}_{mk}^\mathrm{H} \},
\end{align}
(\textit{ii}) follows separating $\mathbf{g}_{mj}$ into its estimate ($\hat{\mathbf{g}}_{mk}$) and estimation error ($\tilde{\mathbf{g}}_{mk}$). The first term is completed by utilizing the results from ( Eq. (4.18),\cite{demir2021foundations}) in which $\hat{\mathbf{g}}_{mj}$  is estimated as
\begin{align}
    \hat{\mathbf{g}}_{mj} = \sqrt{\frac{p_j^\mathrm{p}}{p_k^\mathrm{p}}} \beta_{mj} \beta_{mk}^{-1} \hat{\mathbf{g}}_{mk},
\end{align}
and
\begin{align}
    &\mathbb{E} \{ \hat{\mathbf{g}}_{mj} \hat{\mathbf{g}}_{mj}^\mathrm{H} \hat{\mathbf{g}}_{mk} \hat{\mathbf{g}}_{mk}^\mathrm{H} \} \nonumber \\ &= \frac{p_j^\mathrm{p}}{p_k^\mathrm{p}} \mathbb{E} \{ \beta_{mj} \beta_{mk}^{-1} \hat{\mathbf{g}}_{mk} \hat{\mathbf{g}}_{mk}^\mathrm{H} \beta_{mk}^{-1} \beta_{mj } \hat{\mathbf{g}}_{mk} \hat{\mathbf{g}}_{mk}^\mathrm{H} \} \nonumber \\ &= \frac{p_j^\mathrm{p}}{p_k^\mathrm{p}} \mathbb{E} \{ | \hat{\mathbf{g}}_{mk} \beta_{mj} \beta_{mk}^{-1} \hat{\mathbf{g}}_{mk}^{-1}|^2 \} \nonumber \\ &=
    p_k^\mathrm{p} p_j^\mathrm{p} |\tau_p \beta_{mj} \theta_{mk}^{-1} \beta_{mk} |^2 + p_k^\mathrm{p} \tau_p \big( (\beta_{mj}- \gamma_{mj}) \beta_{mk} \theta_{mk}^{-1} \beta_{mk}   \big),
\end{align}
where $\gamma_{mj}= \beta_{mj} - p_j^\mathrm{p}\tau_p \beta_{mj} \theta_{mk}^{-1} \beta_{mj} $.\\ 
Also, the second term becomes
\begin{align}
    \mathbb{E} \{ \tilde{\mathbf{g}}_{mj} \tilde{\mathbf{g}}_{mj}^\mathrm{H} \hat{\mathbf{g}}_{mk} \hat{\mathbf{g}}_{mk}^\mathrm{H} \} =  p_k^\mathrm{p} \tau_p \gamma_{mj} \beta_{mk} \theta_{mk}^{-1} \beta_{mk}.
\end{align}
By using and adding these two terms together, we obtain \eqref{ujk}.

\add{\section*{Appendix B} \label{APPendix B}
The standard form of GP is defined as follows \cite{chiang2007power}:
\begin{align*}~&\min ~ f_{0}(\textbf {x}),\\&\text {s.t.}\quad f_{i}(\textbf {x}) \!\le \! 1, ~i\!=\! 1,\cdots , m, \\&  \qquad q_{i}(\textbf {x}) \!=\! 1, ~i\!=\! 1,\cdots , c, \end{align*}
where $f_0$ and $f_1$ are posynomial and $q_i$ are monomial functions. Moreover, $x=\{x_1, \cdots x_n\}$ represent the optimization variables. The SINR constraint in \eqref{SINR closed form} can be written as 
Numerator (Desired Signal Power) as:
   \begin{align} \label{No}
   \text(No)_k = N \tau_p p_k^\mathrm{d} p_k^\mathrm{p} \sum_{m=1}^M \beta_{mk}^2 \theta_{mk}^{-1}.
   \end{align}
and denominator (interference plus noise) term as:
\begin{align}
   \text(De)_k = \sum_{m=1}^M \Bigg( N \tau_p \sum_{\substack{j=1 \\ j \ne k}}^{K} p_j^\mathrm{d} p_j^\mathrm{p} \beta_{mj}^2 \theta_{mk}^{-1} + \sum_{j=1}^K p_j^\mathrm{d} \beta_{mj} + \sigma^2 \Bigg).
\end{align}
Thus, we can rewrite the SINR as:
\begin{align}
\mathrm{SINR}_k = \frac{\text(No)_k}{\text(De)_k}.
\end{align}
We check if the numerator and denominator can be written in the posynomial form.
In \eqref{No}, since \(N\), \(\tau_p\), \(\beta_{mk}\), and \(\theta_{mk}\) are constants, we have:
\begin{align}
\text(No)_k = p_k^\mathrm{d} p_k^\mathrm{p} \cdot \text{(constant)}.
\end{align}
This is a monomial, which is compatible with GP.\\
The denominator, \(\text{De}_k\), is composed of several terms:}
\add{\begin{quote}
    \item[1-] \textit{Inter-user interference term:} This is given by \(N \tau_p \sum_{j \ne k} p_j^\mathrm{d} p_j^\mathrm{p} \beta_{mj}^2 \theta_{mk}^{-1}\). It represents a sum over all users \(j \ne k\), where each term is a product of \(p_j^\mathrm{d}\), \(p_j^\mathrm{p}\), and a constant, making it a monomial.
    \item[2-]   \textit{Intra-user interference term:} This is expressed as $\sum_{j=1}^K p_j^\mathrm{d} \beta_{mj}$. Here, each term is a product of $p_j^\mathrm{d}$ and a constant, which also qualifies as a monomial.
    \item[3-]  \textit{Noise term:} Represented by \(\sigma^2\), this is simply a constant.
\end{quote}
As a result, $(\text{De}_k)$ is a sum of these monomials, which collectively form a posynomial. Since posynomials are compatible with GP, this structure aligns well with the requirements for GP-based optimization.}

\bibliographystyle{IEEEtran}
\bibliography{reference}

\end{document}